\documentclass{article}

\usepackage{PRIMEarxiv}

\usepackage{float}
\usepackage[caption = false]{subfig}
\usepackage[export]{adjustbox}
\usepackage{listings}
\usepackage{multicol}
\usepackage{lipsum}

\usepackage{balance}

\usepackage{todonotes}

\usepackage{amsmath,amsfonts}
\usepackage{algorithmic}
\usepackage{graphicx}
\usepackage{textcomp}
\usepackage{xcolor}

\definecolor{codegreen}{rgb}{0,0.6,0}
\definecolor{codegray}{rgb}{0.5,0.5,0.5}
\definecolor{codepurple}{rgb}{0.58,0,0.82}
\definecolor{backcolour}{rgb}{0.95,0.95,0.92}
\lstdefinestyle{mystyle}{
    belowskip=-5 pt,
    backgroundcolor=\color{backcolour},   
    commentstyle=\color{codegreen},
    keywordstyle=\color{magenta},
    numberstyle=\tiny\color{codegray},
    stringstyle=\color{codepurple},
    basicstyle=\ttfamily\footnotesize\linespread{0.8},
    breakatwhitespace=false,         
    breaklines=true,                 
    captionpos=t,                    
    keepspaces=true,                 
    numbers=left,                    
    numbersep=4pt,                  
    showspaces=false,                
    showstringspaces=false,
    showtabs=false,                  
    tabsize=2
}

\usepackage{lipsum} 
\usepackage{soul} 
\title{Integrated Photonic AI Accelerators under Hardware Security Attacks:\\ Impacts and Countermeasures
\thanks{\textit{\underline{Citation}}: 
\textbf{To appear on the procedings of IEEE MWSCAS 2023}} 
}

\author{
  Felipe G\"{o}hring de Magalh\~{a}es, Gabriela Nicolescu \\
  Ecole Polytechnique de Montreal \\
  Montreal, QC \\
  Canada\\
  \texttt{\{felipe.gohring-de-magalhaes, gabriela.nicolescu\}@polymtl.ca} \\
   \And
  Mahdi Nikdast \\
  Colorado State University \\
  Fort Collins, CO \\
  USA\\
  \texttt{mahdi.nikdast@colostate.edu} \\
}

\begin{document}
\maketitle

\begin{abstract}
Integrated photonics based on silicon photonics platform is driving several application domains, from enabling ultra-fast chip-scale communication in high-performance computing systems to energy-efficient optical computation in artificial intelligence (AI) hardware accelerators. Integrating silicon photonics into a system necessitates the adoption of interfaces between the photonic and the electronic subsystems, which are required for buffering data and optical-to-electrical and electrical-to-optical conversions. Consequently, this can lead to new and inevitable security breaches that cannot be fully addressed using hardware security solutions proposed for purely electronic systems. This paper explores different types of attacks profiting from such breaches in integrated photonic neural network accelerators. We show the impact of these attacks on the system performance (i.e., power and phase distributions, which impact accuracy) and possible solutions to counter such attacks. 
\end{abstract}


\section{Introduction}\vspace{-10pt}
    Silicon photonic (SiPh) integrated circuits exploit the fast optical-domain data transmission to realize ultra-high bandwidth communication with low power consumption in high-performance computing systems~\cite{9044439}. SiPh has been deployed in many-core systems to improve their communication infrastructure~\cite{oin_why_2}, and most recently, in artificial intelligence (AI) hardware accelerators to boost both their communication and computation performance~\cite{clements, dl_siph}. 
    
    Integrating photonic and electronic systems (i.e., optoelectronic systems) necessitates the use of opto-electrical interfaces  to deal with signals from different domains. Such integration also necessitates SiPh node configuration (e.g., to adjust signal phase) for routing and computation~\cite{eu_ofc_2016}. For instance, the SiPh-based AI accelerator (SPAA) proposed in \cite{clements} develops a light-speed matrix-multiplication unit whose operation is based on a diagonal decomposition methodology. The matrix defined in the digital domain (i.e., application executing on a processor) needs to be properly mapped to the SPAA. This is done by defining the correct transformation signal phases in each SiPh node, where an electrical controller adjusts the tuning circuitry (e.g., by applying a bias voltage) on each SiPh node to realize the required optical phase. The controller has access to each SiPh node through an interface connection. The same is true for the matrix inputs, mapped as the inputs of the SiPh circuit. Recovering the results in the digital domain is performed by optical-to-electrical conversions and photodetectors.
  
    Systems integrating different technologies (e.g., an optoelectronic system) are more susceptible to malicious attacks~\cite{opt_mitigation3}. In optoelectronic systems, attackers can profit from multiple signal conversions required to exchange data between the two domains and act on the integration interfaces. Furthermore, any photonic device operation is determined by the device's design parameters (e.g., waveguide width) and temperature, where minimal changes can change the device operation. For example, considering the thermal sensitivity of SiPh devices, in the SPAA proposed in \cite{clements}, an attacker (e.g., an IP integrated in the system) can increase its own temperature (e.g., by performing heavy mathematical operations). Consequently, this can impose thermal crosstalk~\cite{Kyatam:20}, creating signal noises by altering the adjusted signal phase shifts in the SiPh devices in proximity. Such phase noises will impact the inferencing accuracy in the SPAA (e.g., by up to 70\% as reported in \cite{9474000,9766130}), degrading the overall system performance. The impact of different attacks can include data leakage and manipulation, service and sleep-denial, irreversible losses, systemic abnormal behaviour, and permanent breakdown~\cite{sergio}.

\begin{figure*}[!t]
    \centering
        \subfloat[SiPh building blocks. MZI operation states: (i) Cross and (ii) Bar. MRR operation states: (iii) OFF resonance and (iv) ON resonance.]{\includegraphics[scale=0.3,valign=t]{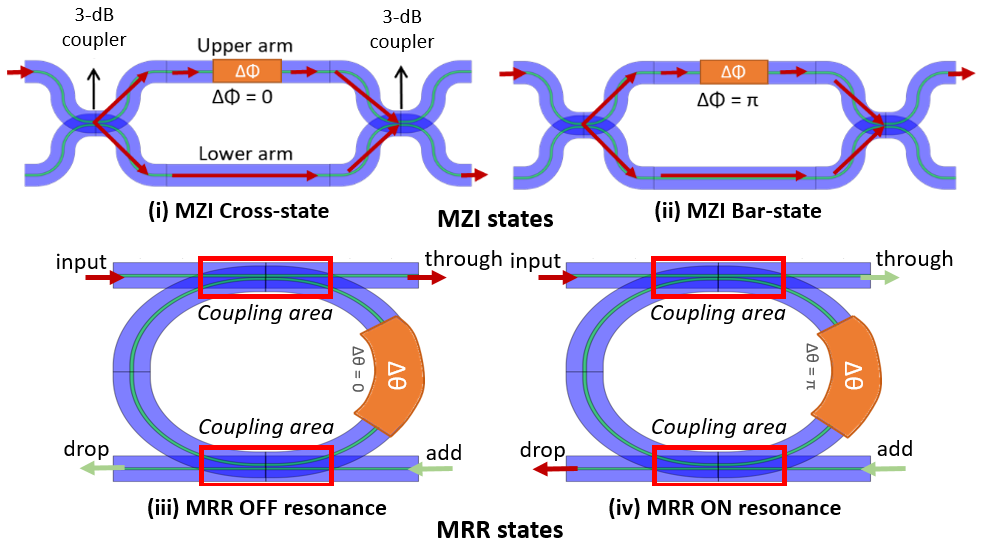}} 
        \hfill \subfloat[Overview of a controller integration with a SPAA. The electrical controller interfaces with the SPAA to configure it (e.g., set optical phases).]{\includegraphics[scale=0.3,valign=t]{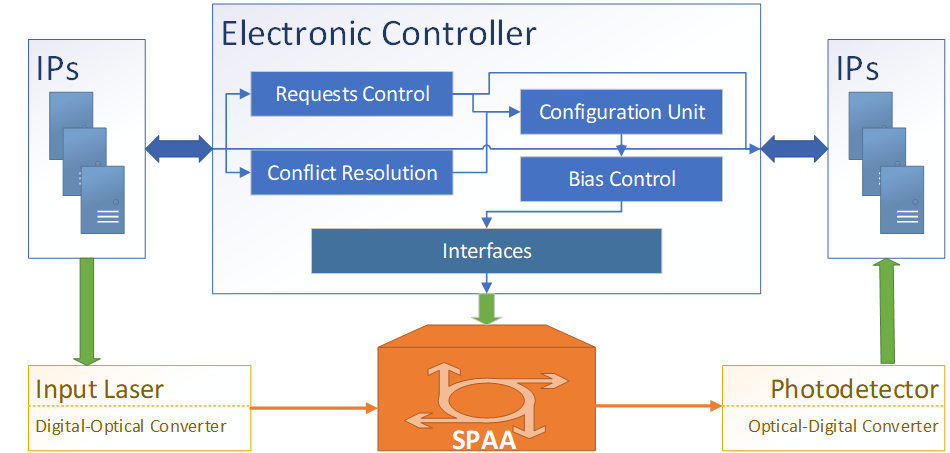}}
        \caption{(a) SiPh building blocks and (b) an overview of controller integration with a SPAA.}\vspace{-10 pt}
    \label{fig:nodes_circuit}
\end{figure*}
    
    Techniques to address electronic hardware Trojan (e-HT) attacks can be explored for optoelectronic systems. Still, most of such techniques are inefficient and will fail to address optoelectronic hardware security concerns~\cite{opt_mitigation}. This is due to the very nature of the breaches in such systems. While e-HTs target IPs and electrical paths, optical HTs (o-HTs) target the optoelectronic interfaces and introduce disturbances to the SPAA. Prior efforts~\cite{9241713,opt_mitigation3} presented solutions to enhance the security of optoelectronic systems by taking into account the characteristics of the optical path, adding an extra layer of security to the design. In this paper, we discuss attacks and their impacts targeting SPAAs and a defence mechanism to detect such attacks. As discussed, SPAAs are prone to  interferences, which can hinder their employment. Hence, tackling security attacks is essential for further advancement and employment of such accelerators. 

\section{Photonic Devices and AI Accelerators}\vspace{-5pt}
    Mach--Zehnder interferometers (MZIs) and microring resonators (MRRs) are widely used as building blocks in different SiPh integrated circuits~\cite{lukas}. Fig. \ref{fig:nodes_circuit}(a) illustrates conventional MZI and MRR designs. An MZI is an interferometric device that includes two 3-dB couplers and optical phase shifters on one or both arms. A phase shifter can be implemented using electro-optic or thermo-optic tuning mechanisms to introduce an optical phase shift on the electric field of optical signals traversing the MZI arms. As a result, the optical signals entering the output coupler can experience destructive or constructive interference (or somewhere in between). For example, in the 2$\times$2 MZI shown in Fig. \ref{fig:nodes_circuit}(a), the phase difference of $\Delta\Phi=$~0 or $\pi$ results in constructive or destructive interference. MZIs are often used in the design of optical switched networks and coherent SPAAs~\cite{Zhou2022,8660861}.
    
    An MRR is a resonating wavelength-selective device. Considering Fig. \ref{fig:nodes_circuit}(a), an optical signal on the input port can be controlled using a biased arm on the ring. The resonant wavelength in an MRR can be adjusted by applying electro-optic or thermo-optic tuning to the MRR. The input signal propagates towards the through port when the ring is \textit{OFF resonance}, and when the MRR is \textit{ON resonance}, the input signal will couple into the ring and goes to the drop port. MRRs are often used in the design of photonic switched networks, modulators, filters, and noncoherent SPAAs~\cite{dl_siph}. For example, the work in \cite{8879624} proposes to use MRRs to create non-linear functions defined by leveraging the impact of  electro/thermo-optic effects in MRRs. 

    SiPh node configuration can be static (passive), defined during design time, or dynamically configured during execution time. Such dynamic reconfigurations can be achieved electronically following the application needs. Fig. \ref{fig:nodes_circuit}(b) illustrates an example of an electronic controller integrated with a SPAA. In this example, each SiPh node in the SPAA is connected to the controller, which can configure the desired operation of the SiPh node. It can interact with the SPAA through electro-optic interfaces, and by using methods such as thermal tuning (e.g., using microheaters) or carrier injection (e.g., through PN junctions). Accordingly, an electronic controller can, for example, configure an MZI by applying a given voltage to increase the temperature in a microheater placed on top of an MZI arm, to change the optical signal phase in the MZI. Taking as an example a coherent SPAA~\cite{clements}, optical phases are adjusted in the underlying SiPh nodes (e.g., MZIs in \cite{clements}) by an electronic controller. Such adjusted phase values represent the weight parameters in the neural network, which can be obtained using software training and decomposition algorithms~\cite{clements}. 
    
    Using both MZIs and MRRs can help design powerful photonic computing architectures capable of performing complex operations. In general, SPAAs use a combination of power intensity and signal phases on the outputs to mimic different operations. For example, a photonic processor using both MZIs and MRRs can be used to perform Fourier transforms, which are essential for applications such as signal processing, image recognition, and data compression~\cite{fourier}. Coherent SPAAs based on MZIs, considered as an example in this paper, can be designed by performing coherent multiplication between an input vector $I_n$ (i.e., input optical signals modulated with input features) and a weight matrix $W$ (i.e., defined by adjusting phase settings on different MZIs), to obtain an output vector $O_n$. The linear multipliers can be represented using two unitary multipliers and a diagonal matrix, which are obtained using singular value decomposition (SVD) \cite{clements}. The multipliers and the diagonal matrix can be realized using a network of interconnected MZIs with different typologies. In~\cite{mcgill_acc, mit_acc}, an $N\times N$ neural network is realized using a series of special unitary (SU) groups. Each SU is built using an MZI with two phase controllers, denoted SU(2). Based on SU(2), larger SU($N$) groups are built and the entire configurable neural network can be created, in which $N$ represents the number of inputs and outputs of the neural network  (i.e., $\forall~ I_n \rightarrow I_n \times W = O_n$).

\section{SPAAs under Hardware Security Attacks}\vspace{-5pt}
    
    \begin{figure*}[!t]
        \centering
            \subfloat[Output readings with and without variations.]{\includegraphics[scale=0.2,valign=t]{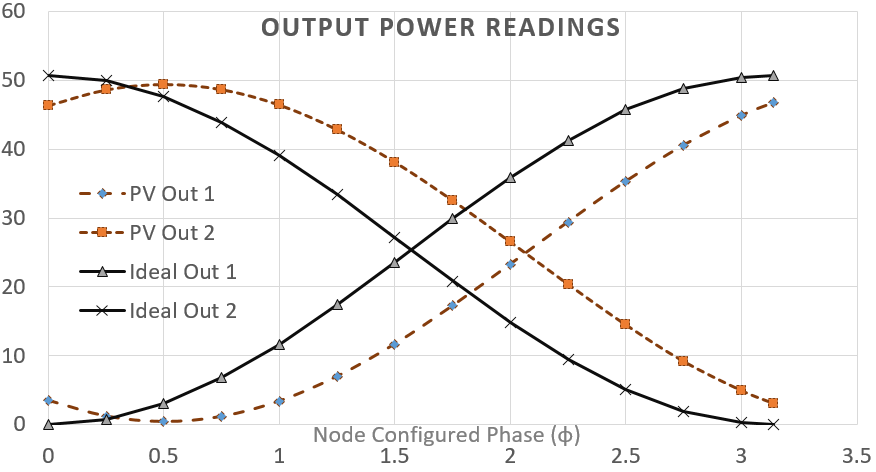}}
            \hfill \subfloat[Validation environment presenting the main blocks composing the testbed.]{\includegraphics[scale=0.17,valign=t]{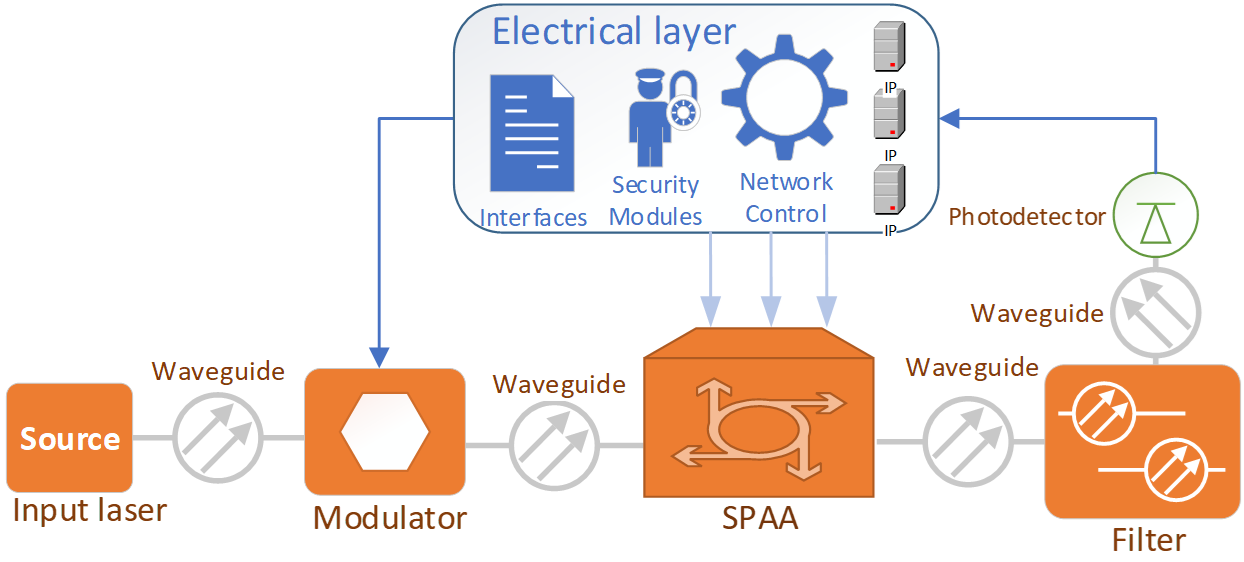}}
            \hfill \subfloat[Unitary matrix of a photonic complex-valued neural network (PCNN) \cite{PCNN}.]{\includegraphics[scale=0.16,valign=t]{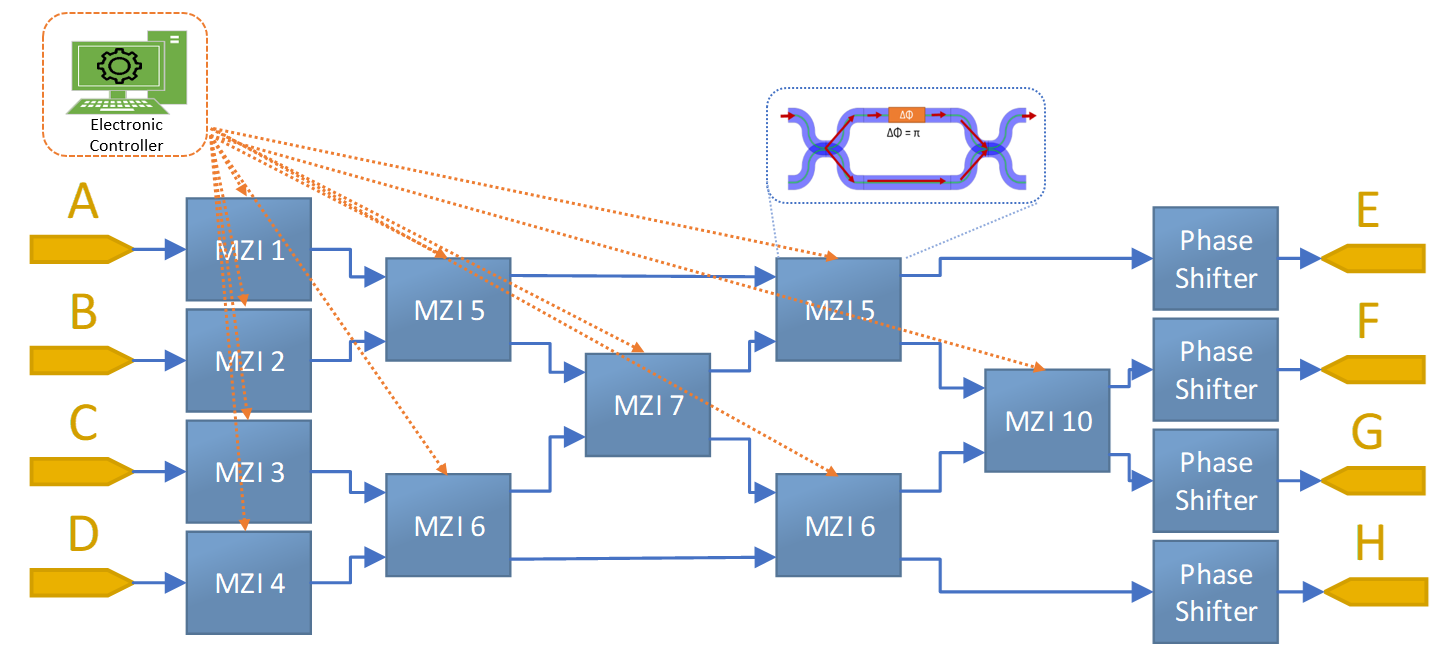}}
             \caption{(a) Optical linear multiplier example architecture and (b) validation environment (c) Output readings considering variations and not.}\vspace{-10 pt}
        \label{fig:overview}
    \end{figure*}
    
    Modern systems are prone to attacks due to a variety of vulnerabilities. These range from software bugs leading to undesired behaviour to maliciously inserted malware in a system. Pure electronic and optoelectronic systems share many of such vulnerabilities. One type of an attack that has the potential to severely affect a system and is hard to detect is the Hardware Trojan (HT). HTs act to steal information, degrade the performance, or even destroy the attacked system. Such type of an attack is hard to detect as it can remain in an ``idle state'' until some action triggers its execution \cite{7543420}. Although sharing many vulnerabilities, optoelectronic systems have unique breaches and affected areas, needing special security treatment. 

    Denial-of-service (DoS) attacks can be injected by HTs. For instance, a sinkhole attack is a type of an attack in which one node actively collects data from a network path. This can be performed by an invaded node that redirects a transmission in the network to a different direction/destination. Another type of the same attack works by making false requests for neighbouring routers in order to receive transmissions addressed to other nodes. A flooding attack is a different type of DoS attack where an invaded IP/node affects the transmissions in the network by injecting noise/adversarial signals. This type of an attack aims at depriving the connected nodes from properly using the transmission path, as transmissions with too much noise are occurring. In a black-hole attack, one or more nodes are tweaked in such a way that the transmission(s) will target dead ends, leading to packet drops. 

    \begin{figure*}[!t]
        \centering
            \subfloat{\includegraphics[scale=0.23,valign=t]{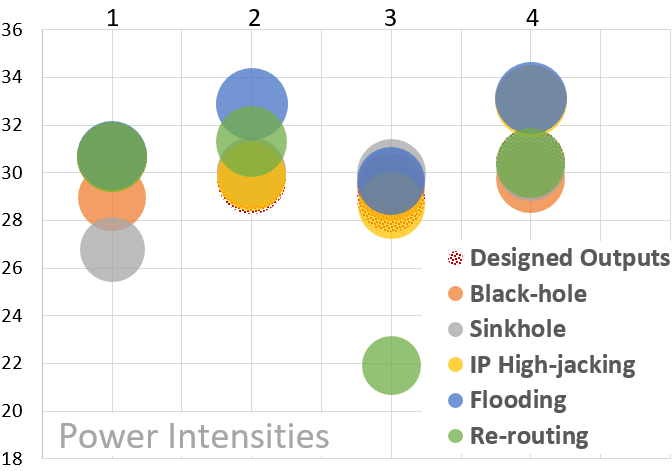}} 
            \hfill \subfloat{\includegraphics[scale=0.2,valign=t]{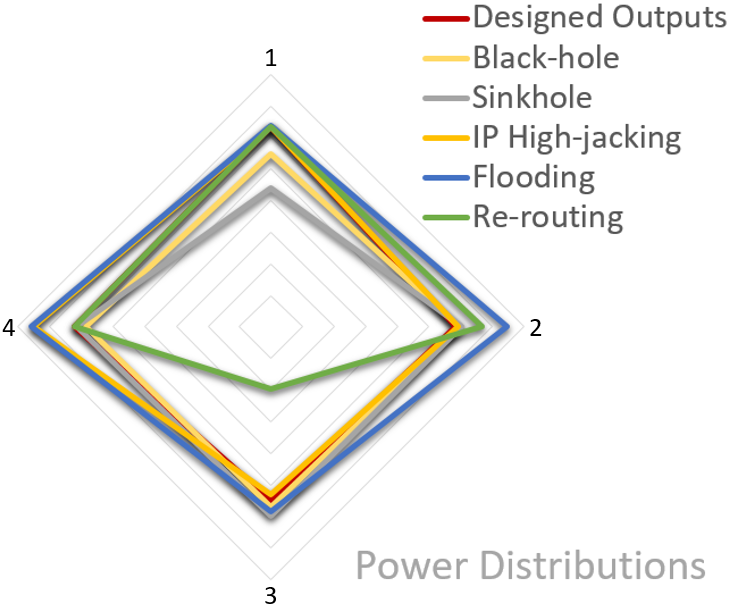}}
            \hfill \subfloat{\includegraphics[scale=0.23,valign=t]{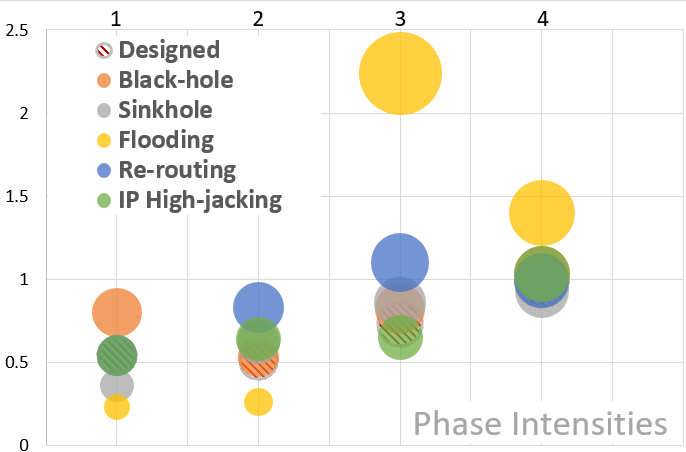}}
            \hfill \subfloat{\includegraphics[scale=0.2,valign=t]{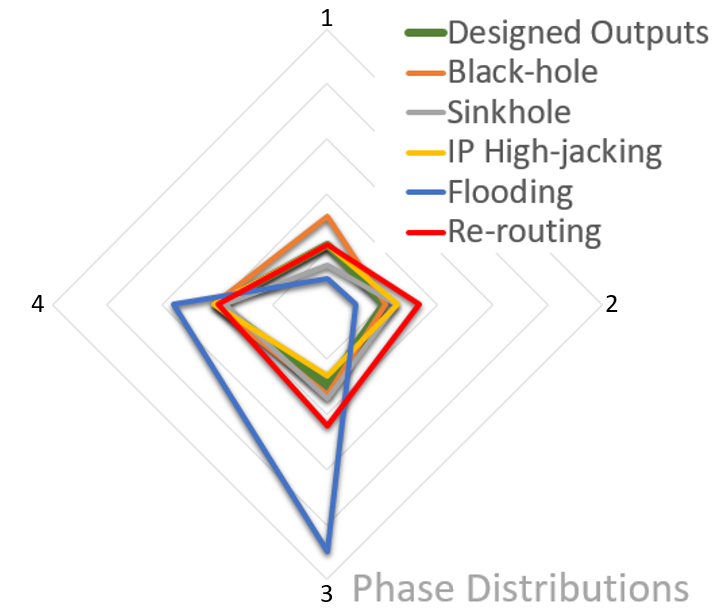}}
            \caption{Output readings with and without perturbances, showing the impact on the outputs for each type of attack. The x-axis presents the outputs. On the y-axis, power is presented in dBm and phase is in radians.}\vspace{-10 pt}
        \label{fig:outputs}
    \end{figure*}
    
    SiPh devices are sensitive to different variations, including fabrication-process (PV) and thermal variations. Prior studies showed that such variations will create optical phase noises in SPAAs, which can cause up to a 70\% reduction in the inferencing accuracy of the network~\cite{9474000,9766130}. SPAAs are vulnerable to attacks that may leverage from HTs to inject noise into channels (e.g., cause changes in adjusted phases). Malicious IPs (e.g., with infected software components) can also increase the temperature of neighbouring nodes, which can result in variances in the physical characteristics of SiPh nodes, and ultimately, lead to incorrect operations. Another study observed a decrease in the accuracy of a SPAA by 5\% when the temperature was increased from 25°C to 50°C. Moreover, it was demonstrated that the presence of optical phase noise can lead to a reduction in the network's accuracy, with an average error rate of 3.7\%~\cite{9474000}.
        
    Although solutions are proposed for electronic HTs~\cite{FREY201715}, not all of them are directly applicable to optoelectronic systems. For optoelectronic systems, efforts are made to ensure the security of designs by relying on the characteristics of the circuit~\cite{ZHANG201884}. Some solutions have been presented to enhance the security of optoelectronic systems, taking advantage of different aspects of SPAAs, such as PVs~\cite{9465434}. In~\cite{opt_mitigation}, the focus is on tampering and snooping attacks during the thermal sensing in photonic networks-on-chip (PNoCs). A detection scheme was presented based on spiking neural networks (SNNs) trained to learn network patterns. By using thermal sensing and SNN feedback, the overall system security against tampering and snooping attacks is enhanced. The work in \cite{opt_mitigation2} presented a framework that utilizes PV-based authentication signatures. A new layer is proposed, integrated in the network gateway interface, which acts to encrypt transmitted data and detect attacks in runtime. In~\cite{opt_mitigation3}, a technique was proposed based on the integration of the network controller with the detection mechanism while focusing on tackling HT attacks. SecONet was presented in \cite{9241713} aiming at securing a SiPh-based network against eavesdropping, spoofing, replay, and message-removal attacks. This is achieved by using a combination of speculative and pre-computation, where a security layer and key-generating units are added to the network hardware stack. Nevertheless, the aforementioned efforts are all limited by different factors, such as strong dependence on the electronic controller, using pre-trained neural networks, or complex and error-prone off-line evaluations, such as experimentally measuring deviations. Also, no prior work has studied SPAAs and proposed countermeasures under security attacks.  
    
    In this paper, we explore a non-intrusive method relying on side-channel analysis. Side-channel analysis is a technique used to detect perturbations by measuring the output readings of the SPAA during operation. To do that, a hardware module is integrated with the electronic layer of the system and interacts with the SPAA to collect information of the transmitted signals. Attacks, such as black holes for instance, will interfere with the signal transmissions in the SPAA, which will lead to noise being inserted on the transmitted signals. The noise difference can be used though as a mechanism to detect perturbations (e.g., malicious interactions or attacks) in the SPAA.\vspace{-5pt}

\section{Countermeasuring the Breaches}\vspace{-5pt}
    SPAAs are susceptible to various types of attacks that can impact their accuracy directly~\cite{9474000}. As demonstrated by ~\cite{opt_mitigation, opt_mitigation3}, perturbation detection in a SPAA is feasible by gathering underlying transmission information and analyzing transmission patterns. Computer-aided machine learning algorithms or modified network electronic controllers can perform this task. However, these methods necessitate either the transfer of data to perform off-line calculations or the modification of integrated modules, which is rarely feasible. In contrast, our approach here relies on side-channel analysis to detect variations in the transmitted signals. This approach enables real-time operational analysis while being
    agnostic to the SPAA architecture. 
    
    We introduce a two-step technique for attack detection in SPAAs. In the first step, during the initialization phase, baseline operation scenarios are established, including input laser(s) information and SiPh node configuration parameters. Reference output values are stored for later use. In the second step, during execution time, the same baseline scenarios are repeated, and the new output values are compared with the stored ones. If the difference between the readings exceeds a configured threshold, an alarm is triggered, and the system is flagged as suspicious. This paper focuses solely on the intrusion detection system (IDS) and does not address finding the point of breach. High-level entities in the system, such as the operating system or network controller, can handle the alarm. The motivation behind this technique stems from the low tolerance of SiPh nodes to variations. Fig.~\ref{fig:overview}(a) illustrates the impact of variances on an MZI node, similar to the one shown in Fig.\ref{fig:nodes_circuit}(a). As it can be seen, variations in the node cause deviations in the readings from their expected values. Such deviations impact the SPAA system performance, as shown by~\cite{9474000}. 

    Our technique is designed to detect runtime perturbations and inform the system about the issue, allowing for any modifications to the SPAA to be detected by our module. More importantly, no changes to the internal hardware modules (such as the network controller) or the transfer of data to external computing hosts (such as for machine learning algorithms) are required. Instead, our module can be seamlessly integrated into the underlying system and act as an integrated module.

\section{Results and Discussions}\vspace{-5pt}
    To validate our technique, we used Ansys Lumerical Interconnect to simulate dfferent SPAAs~\cite{clements,mit_acc,mcgill_acc,PCNN}. The simulation environment is illustrated in Fig.~\ref{fig:overview}(b), and all electrical modules were described using Python for seamless integration with SiPh models. We employed different SPAA architectures in our study~\cite{clements,mit_acc,mcgill_acc}, including the one presented in Fig.~\ref{fig:overview}(c), which is used in~\cite{PCNN} as part of a photonic complex-valued neural network (PCNN). 

    To provide insight into the impact of different types of attacks, such as black-hole, sinkhole, re-routing, IP high-jacking, and flooding, we present the outputs under normal conditions and under attacks in Fig.~\ref{fig:outputs}, for the architecture presented in Fig.~\ref{fig:overview}(c). The figures show the output power intensities and phases distributions. Under normal operation, all the readings should be identical or as close as possible to the designed ones. As it is possible to see, under different attacks, many variations on the readings occur, which lead to a decrease in the network accuracy, as previously reported in the literature\cite{onn_under}. By using these readings, our side-channel IDS can easily detect attacks and trigger an alarm. Despite its simplicity, our approach is effective because of the low variance required on the SiPh nodes to impact the outputs, as shown by our experimental results --- see Fig.~\ref{fig:outputs}.

    Our approach offers an efficient and non-intrusive technique for detecting perturbations in SPAAs. The proposed method only requires baseline operation scenarios to be established during the initialization phase, and during execution time, the system compares the new output values with the stored reference output values. If the difference between the readings is greater than a specified threshold, an alarm is triggered. Our approach does not require any modification on the internal hardware modules or the transfer of data to outside computing hosts. With this approach, we can successfully detect any perturbation on the SPAA without incurring additional overhead.
    
\section{Conclusion}\vspace{-5pt}
Silicon-photonic-based AI accelerators (SPAAs) require different opto-electrical interfaces for proper function (e.g., data storage). But using such interfaces make SPAAs vulnerable to various types of attacks. This paper presented a study of different attacks that exploit such vulnerabilities and shows their impact on the system performance. We also proposed possible countermeasures to mitigate these attacks. Our technique is validated using circuit models for SPAA simulation, and we demonstrated its ability to detect different attacks. We believe that this technique offers a valuable tool for ensuring the safety of emerging SPAAs.

\section*{Acknowledgement}\vspace{-5pt}
This work was supported in part by the National Science Foundation (NSF) under grant number CNS-2046226.

\balance
\bibliographystyle{unsrt}
\bibliography{Arxiv_MWSCAS}
\end{document}